\documentclass[aps,prb,preprint]{revtex4}
\usepackage{epsfig,amssymb,amsfonts}
\usepackage{graphicx}
\input{epsf}
\begin{document}
\draft

\title{Thermal emission from three-dimensional arrays of gold nanoparticles}

\author{Vassilios Yannopapas}
\email{vyannop@upatras.gr}
\address{Department of Materials Science, School of Natural Sciences, \\
         University of Patras, GR-26504 Patras, Greece}

\begin{abstract}
We study the blackbody spectrum from slabs of three-dimensional
metallodielectric photonic crystals consisting of gold
nanoparticles using an ab initio multiple-scattering method. The
spectra are calculated for different photonic-crystal slab
thicknesses, particle radii and hosting materials. We find in
particular that such crystals exhibit a broadband emission
spectrum above a specific cutoff frequency with emissivity of
about 90\%. The studied photonic crystals can be used as efficient
selective emitters and can therefore find application in
thermophotovoltaics and sensing.
\end{abstract}
 \maketitle
\bibliographystyle{apsrev}
\narrowtext

The main feature of photonic crystals is the ability to tailor the
photon density of states and this way control the spontaneous
emission of light, aiming at the realization of new optoelectronic
devices. In this context, there has been considerable effort to
design and fabricate photonic crystals which allow for control of
thermal emission of light, i.e. thermally driven spontaneous
emission, promising applications in imaging, sensing and most
importantly, in thermophotovoltaics (TPV).
\cite{cornelius,lin1,pralle,fleming,lin2,celanovic,chen,enoch,florescu}
Control of thermal emission can also be achieved by means of
microstructured engineering on silicon
\cite{sai,maruyama,greffet1,greffet2} or metal surfaces.
\cite{kusunoki} Depending on the type of application, photonic
crystals and structured surfaces can act as narrow- or wide-band,
directional or isotropic thermal emitters. For example, in TPV
applications \cite{zenker} a quasi-monochromatic emission is
preferable whilst in radiation cooling \cite{hoyt} a broad
emission spectrum is desired. In this work we investigate the
emission properties of three-dimensional metallodielectric
photonic crystals consisting of gold nanoparticles. We find, in
particular, that the emission spectrum of these crystals can be
such that photons are emitted in all directions only when their
energies lie above a specific cutoff frequency, with emissivity as
large as 90\%.

Photonic crystals of spherical scatterers have been theoretically
studied using multiple scattering theory \cite{skm92,comphy} which
is ideally suited for the calculation of the transmission,
reflection and absorption coefficients of an electromagnetic (EM)
wave incident on a composite slab consisting of a number of planes
of non-overlapping particles with the same two-dimensional (2D)
periodicity. For each plane of particles, the method calculates
the full multipole expansion of the total multiply scattered wave
field and deduces the corresponding transmission and reflection
matrices in the plane-wave basis.  The transmission and reflection
matrices of the composite slab are evaluated from those of the
constituent layers. Having calculated these matrices one can
evaluate the transmittance ${\mathcal{T}}(\omega,\theta, \phi)$,
reflectance ${\mathcal{R}}(\omega,\theta, \phi)$, and from those
two, the absorbance ${\mathcal{A}}(\omega,\theta, \phi)$ of the
composite slab as functions of the incident photon energy $\hbar
\omega$ and incident angles $\theta$ and $\phi$. Transmittance and
reflectance are defined as the ratio of the transmitted,
respectively the reflected, energy flux to the energy flux
associated with the incident wave. The method applies equally well
to non-absorbing systems and to absorbing ones. In terms of speed,
convergence and accuracy, the multiple scattering method is the
best method to treat photonic structures of spherical particles.
The emittance ${\mathcal{E}}(\omega,\theta, \phi)$ is calculated
indirectly by application of Kirchoff's law,
\cite{rytov,vesperinas} i.e. from
\begin{equation}
{\mathcal{E}}(\omega,\theta, \phi)={\mathcal{A}}(\omega,\theta,
\phi)= 1-{\mathcal{R}}(\omega,\theta,
\phi)-{\mathcal{T}}(\omega,\theta, \phi) \label{eq:kirchoff}
\end{equation}
Note that recent direct calculations of the thermal emission from
photonic crystals have verified the validity of Kirchoff's law for
the case of photonic crystals. \cite{luo}

In this work we deal with gold spheres with radius of a few
nanometres. Traditionally, the gold spheres are treated as plasma
spheres whose dielectric function is given by Drude's formula
\begin{equation}
\epsilon_{p}(\omega)= 1-\frac{\omega_{p}^{2}}
{\omega(\omega+\mathrm{i}\tau^{-1})} \label{eq:drude}
\end{equation}
where $\omega_{p}$ stands for the bulk plasma frequency of the
metal and $\tau$ is the relaxation time of the conduction-band
electrons. The dielectric function
$\epsilon(\omega)=\epsilon_{1}(\omega)+ \mathrm{i}\
\epsilon_{2}(\omega)$ for bulk gold has been determined
experimentally by Johnson and Christy \cite{johnson} over the
range from $\hbar\omega=0.64$~eV to $\hbar\omega=6.60$ eV. Taking
$\hbar\tau^{-1}=\hbar\tau_{b}^{-1}=0.027$~eV and
$\hbar\omega_{p}=8.99$~eV \cite{wind,bobbert} Eq.~(\ref{eq:drude})
reproduces satisfactorily the experimentally determined
$\epsilon_{1}(\omega)$ over the whole of the above frequency
range, and the experimentally determined $\epsilon_{2}(\omega)$ at
low frequencies. $\tau_{b}$ refers to the collision time of
conduction-band electrons in  the bulk metal; due to the
finiteness of a nanosphere, the collision time $\tau$ of an
electron in a sphere of radius $S$ is decreased due to scattering
at the boundaries of the sphere. It is given by
\begin{equation}
\tau^{-1}=\tau_{b}^{-1}+v_{F}S^{-1} \label{eq:tau}
\end{equation}
where $v_{F}$ stands for the average velocity of an electron at
the Fermi surface. For gold $\hbar v_{F}= 0.903$~eV$\cdot$nm,
\cite{norman,abeles} so that for a sphere of radius $S=5$~nm,
using $\hbar\tau_{b}^{-1}=0.027$~eV, we obtain
$\hbar\tau^{-1}=0.21$~eV. Therefore, a more realistic description
of the dielectric function of a gold sphere would be
\cite{norman,abeles}
\begin{equation}
\epsilon_{S}(\omega)=\epsilon(\omega)+\frac{\omega_{p}^{2}}
{\omega(\omega+\mathrm{i}\ \tau_{b}^{-1})}- \frac{\omega_{p}^{2}}
{\omega(\omega+\mathrm{i}\ \tau^{-1})} \label{eq:ecorr}
\end{equation}
where $\epsilon(\omega)$ is the experimentally determined
dielectric function of bulk gold. We note that
$\epsilon_{S}(\omega)$ as given above, is different from the
approximation to the dielectric function of the gold sphere
obtained from Eq.~(\ref{eq:drude}) with $\hbar\omega_{p}=8.99$~eV
and $\hbar\tau^{-1}=0.21$~eV only because the imaginary part of
the dielectric function is not represented well by
Eq.~(\ref{eq:drude}) for $\hbar\omega>1.5$~eV. Taking into account
the finite size of a metal nanoparticle for the determination of
the actual dielectric function, i.e. Eq.~(\ref{eq:ecorr}), has
successfully reproduced experimentally obtained light absorption
and scattering spectra of monolayers of gold nanoparticles.
\cite{stef91a}

\begin{figure}[htb]
\centerline{\includegraphics[width=8cm]{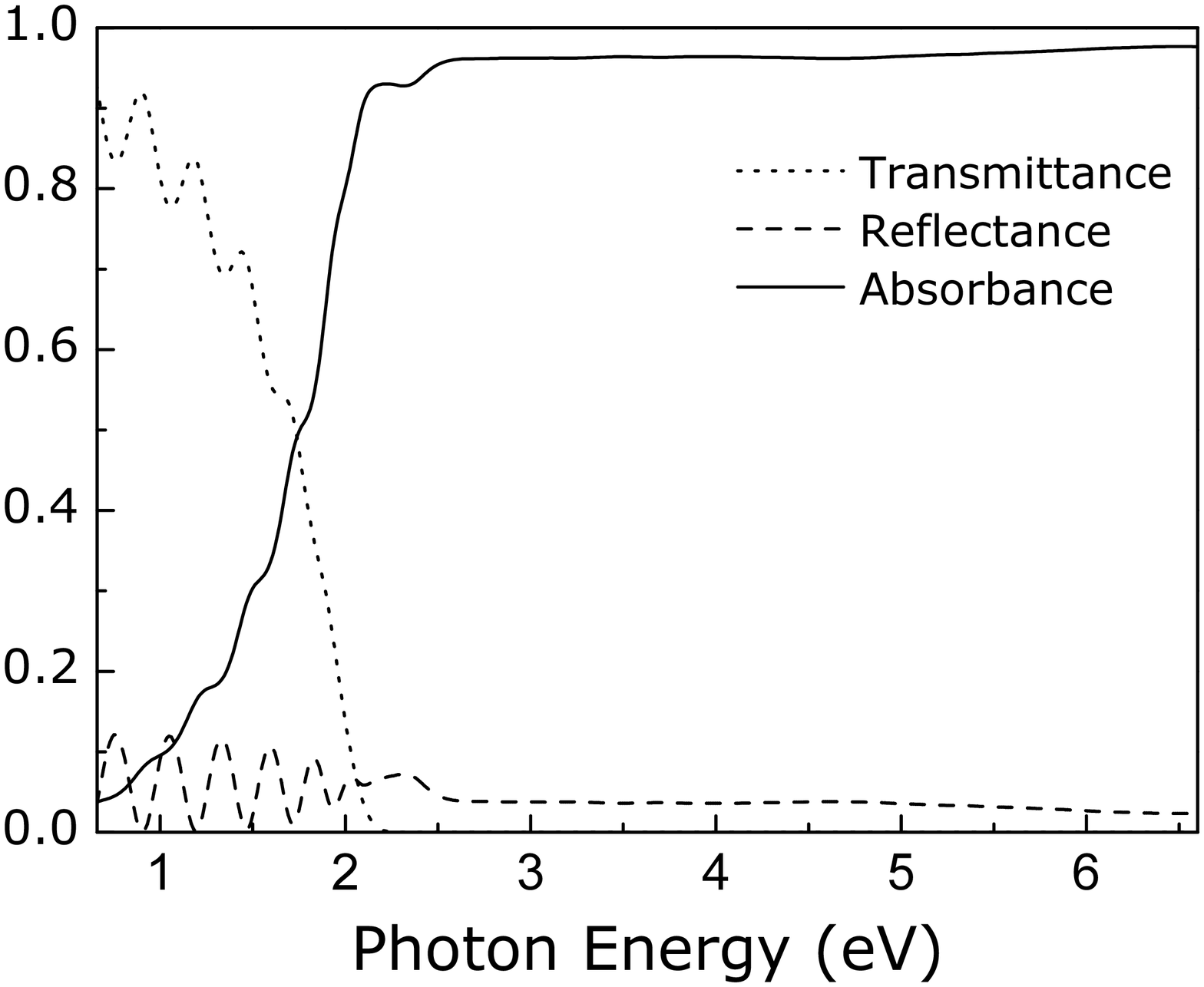}}
\caption{Transmittance, reflectance, and absorbance of light
incident normally on a 128-layers thick slab of a fcc crystal
consisting of gold spheres ($S=5$~nm) in air ($\epsilon_{h}=1$),
with $f=0.3$.} \label{fig1}
\end{figure}

To begin with, we consider an fcc photonic crystal whose lattice
sites are occupied by gold nanospheres of radius $S=5$~nm. The
lattice constant of the crystal is $a=19.11$~nm corresponding to a
volume filling fraction occupied by the spheres, $f=0.3$. The
spheres are supposed to be suspended in air. The crystal is viewed
as a succession of planes of spheres (layers) parallel to the
(001) surface of fcc. In Fig.~\ref{fig1} we show the
transmittance, reflectance and absorbance vs energy for light
incident normally on a finite slab of the crystal consisting of
128 layers of spheres. We observe that for energies above $2.2$~eV
almost all light (92\%-98\%) that is incident on the crystal slab
is absorbed from the gold spheres. Therefore, from Kirchoff's law,
i.e. ${\mathcal{E}}(\omega,\theta,
\phi)={\mathcal{A}}(\omega,\theta, \phi)$, we can infer that the
blackbody radiation is strong above this energy. Indeed, the
blackbody radiation intensity of the photonic crystal is given by
\begin{equation}
I_{PC}(\omega, \theta, \phi, T)=\mathcal{E}(\omega,\theta, \phi)
I_{BB}(\omega, T) \label{eq:pc_intensity}
\end{equation}
where $I_{BB}(\omega, \theta, \phi, T)$ is the blackbody radiation
intensity (Planck distribution)
\begin{equation}
I_{BB}(\omega, T)=\frac{\hbar \omega^{3}} {4 \pi^{3} c^{2}} \
\frac{1} {\exp(\hbar \omega /k_{B} T) -1}. \label{eq:planck}
\end{equation}
$k_{B}$ is the Boltzmann constant and $c$ is the speed of light in
vacuum. In a metal particle of nanoscale radius, there are two
principal sources of light emission/ absorption: the first one is
the dipole oscillations due to the surface plasmon resonances and
the second is the interband transitions that take place in bulk
gold contributing to its dielectric function. In terms of the
first source of emission, the crystal can be viewed as a lattice
of interacting oscillating dipoles where each one of the dipoles
emits light around the surface plasma resonance frequency
$\omega_{SP}=\omega_{p}/\sqrt{3}$. The interaction between the
dipoles leads to the creation of a band of resonances around
$\omega_{SP}$. In two-dimensional systems, this band of resonances
is manifested as a strong absorption/ emission peak \cite{stef91b}
whilst in three-dimensional systems is manifested as two
absorption/ emission peaks (corresponding to two bands of
resonances) with a reflectivity peak (stemming from a photonic
band gap) lying between them. \cite{yanno99} In Fig.~\ref{fig1} we
cannot clearly identify surface plasmon peaks as these must have
been submerged into the absorption edge due to interband
transitions. In any case, both contributions (plasmon+interband)
lead to a strong absorption/ emission plateau above 2.2~eV.

\begin{figure}[htb]
\centerline{\includegraphics[width=8cm]{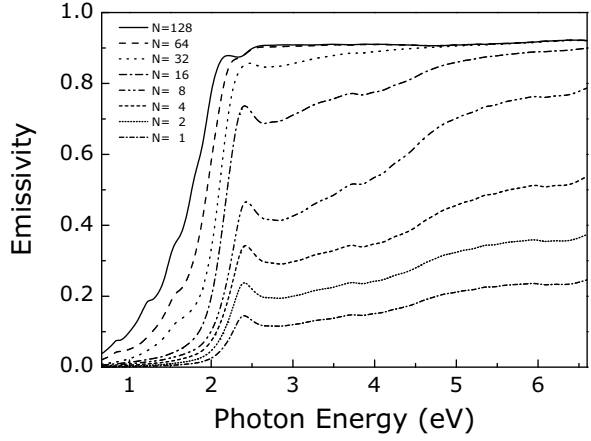}}
 \caption{SH emissivity for different numbers of layers (slab thicknesses)
of the photonic crystal described in Fig.~\ref{fig1}.}
\label{fig2}
\end{figure}

As the system considered here can potentially found application in
TPV devices we are interested in the spectral hemispherical (SH)
radiative properties of the photonic crystal. More specifically,
we are interested in the SH emissivity $s(\omega)$ which is the
ratio of the SH emissive power of a photonic-crystal slab to the
SH emissive power of a perfect blackbody at the same temperature
$T$. For a slab infinitely extended in two dimensions:
\begin{eqnarray}
s(\omega)&=&\frac { \int_{0}^{2 \pi} d\phi \int_{0}^{\pi/2}
d\theta I_{PC}(\omega, \theta, \phi, T) \cos \theta \sin \theta }
{ \int_{0}^{2 \pi} d\phi \int_{0}^{\pi/2}
d\theta I_{BB}(\omega, T) \cos \theta \sin \theta } \nonumber \\
&=& \frac{1}{\pi} \int_{0}^{2 \pi} d\phi \int_{0}^{\pi/2} d\theta
\mathcal{E}(\omega,\theta, \phi) \cos \theta \sin \theta
\label{eq:emissivity}
\end{eqnarray}
Note that $\mathcal{E}(\omega,\theta, \phi)$ in the above equation
is the arithmetical average of both polarization modes. In
Fig.~\ref{fig2} we show the SH emissivity from slabs of the
photonic crystal of Fig.~\ref{fig1} for different slab thicknesses
(number of layers). It is evident that as the number of layers
increases the SH emissivity increases accordingly until it reaches
a saturation plateau for slabs of 64 layers and above. Indeed, for
energies above 2.2~eV the emissivity curves for 64 and 128 layers
are practically the same. This fact implies that, if we measure
the emission from one of the surfaces of a 128 layers-thick slab,
then the emitted radiation must be coming from the 64 outmost
layers (relative to a given surface) whilst radiation coming from
the 64 innermost layers is almost totally absorbed from the 64
outmost layers by the time it reaches the surface of the slab. The
most important finding of Fig.~\ref{fig2} is the fact that the
photonic crystal behaves more or less as gray body (emissivity
ranging from 88\% to 92\%) for energies above 2.2~eV while, at the
same time, emits small amounts of radiation for energies below
this cutoff. A similar emission spectrum is observed for the
Salisbury screen \cite{greffet2} but with significantly lower
values of the SH emissivity.

\begin{figure}[htb]
\centerline{\includegraphics[width=8cm]{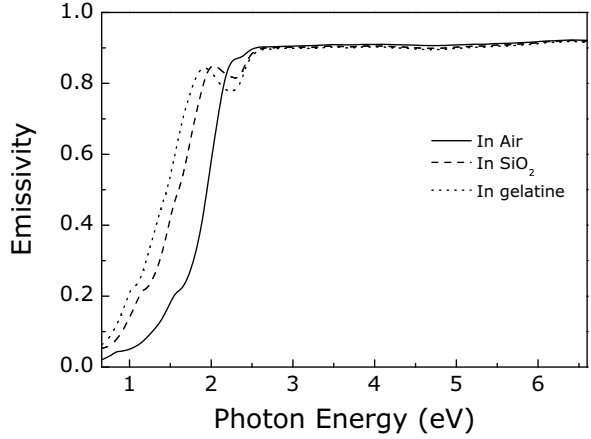}}
 \caption{SH emissivity of a 64-layer thick slab of an fcc crystal
of gold spheres ($S=5$~nm) in air ($\epsilon_{h}=1$ - solid line),
silica ($\epsilon_{h}=1.88$ - dashed line) and gelatine
($\epsilon_{h}=2.37$ - dotted line), with $f=0.3$. We have assumed
that the host medium surrounding the spheres also covers the whole
space.} \label{fig3}
\end{figure}

So far, we have assumed that the gold nanospheres are suspended in
air. In order to provide a manufacturable structure it is
necessary to examine the case where the spheres are embedded in a
host material of dielectric function $\epsilon_{h}$. In
Fig.~\ref{fig3} we show the SH emissivity from a 64-layer slab of
an fcc photonic crystal of gold nanospheres with $f=0.3$ where the
spheres are surrounded by air ($\epsilon_{h}=1$), silica
($\epsilon_{h}=1.88$) and gelatine ($\epsilon_{h}=2.37$). The
introduction of a host material does not change the picture
drastically except that it lowers the emission cutoff energy. This
is due to the lowering of the surface plasmon frequency according
to $\omega_{SP}=\omega_{p}/\sqrt{1+2\epsilon_{h}}$ when a
nanosphere is placed in a host medium $\epsilon_{h}$, which
results in strong thermal emission at lower energies. As a result
of the energy shift of the surface plasmon resonance, the
corresponding emission peak moves away from the interband emission
region and a distinct local minimum appears. This is evident from
the emissivity spectra of Fig.~\ref{fig3} for silica and gelatine
host materials. The shift of the surface plasmon energy with
respect to the host material $\epsilon_{h}$ allows for tuning of
the cutoff energy so that it coincides with the band gap of the
photodiode of a TPV device.

\begin{figure}[htb]
\centerline{\includegraphics[width=8cm]{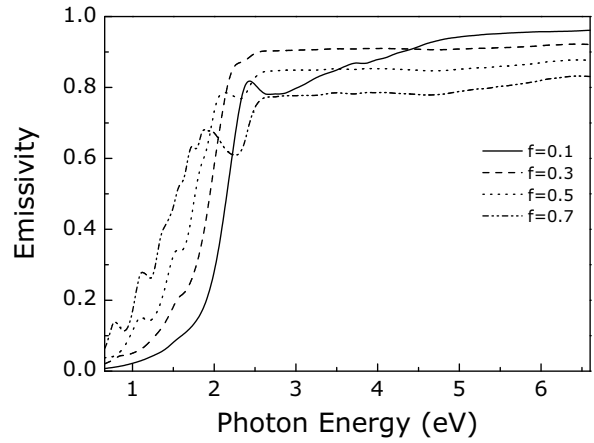}}
 \caption{SH emissivity of a 64-layer thick slab of an fcc crystal
of gold spheres ($S=5$~nm) in air, for different volume filling
fractions $f$ (fixed lattice constant and different sphere
radii).} \label{fig4}
\end{figure}

Next we study the effect of the volume filling fraction $f$ on the
emission properties of the photonic structure under study. In
Fig.~\ref{fig4} we show the SH emissivity spectra for photonic
crystals of different values of $f$. We have kept the lattice
constant the same and changed the sphere radii accordingly. Note
that for each value of $f$ we obtain a different value of $\hbar
\tau^{-1}$ from Eq.~(\ref{eq:tau}). We observe that the maximum
emissivity in the region above the cutoff energy is achieved for
$f=0.3$. Since our calculations show that the SH reflectivity
assumes higher values for $f=0.5, 0.7$, the radiation emitted from
the innermost layers is reflected back and reabsorbed before it
reaches the surface of the slab.

Finally we address the issue of spatial order/ disorder of the
photonic crystal under investigation. As it has been both
theoretically \cite{cpa,oqe} and experimentally \cite{velikov}
shown, the presence of disorder does not changes practically the
absorption/ emission properties of three-dimensional photonic
crystals of metal nanoparticles since these quantities mostly
depend on single sphere properties such as surface plasmon
resonances, and less on the particular spatial arrangement of the
spheres. Note that this not true for one- or two-dimensional
structures, e.g. a linear chain or a monolayer of nanospheres,
where the presence of disorder changes dramatically the
absorption/ emission spectra. \cite{oqe} So, since we deal with a
three-dimensional structure, a disordered crystal might be easier
and cheaper to fabricate.

In summary, we have shown that a metallodielectric crystal
consisting of gold nanospheres can act as a 90\% gray body for
energies above a cutoff. The cutoff frequency can be tuned by
appropriate choice of the material hosting the nanospheres.
Optimal emissivity is obtained for moderate volume filling
fractions and slab thicknesses.

 \section*{ACKNOWLEDGEMENTS}
This work was supported by the `Karatheodory' research fund of
University of Patras.

\small
\normalsize

\small
\normalsize

\small
\normalsize

\small

\end{document}